\documentclass[english,twoside,a4paper,10pt]{article}

\usepackage[latin1]{inputenc}
\usepackage[T1]{fontenc}
\usepackage{amsmath}
\usepackage{amsfonts}
\usepackage{graphicx}
\usepackage{subfigure}
\usepackage{a4wide}
\usepackage{amssymb}
\usepackage{fancyhdr}
\usepackage{mathrsfs}
\usepackage{amssymb}
\usepackage[toc,page]{appendix}
\usepackage{xcolor}
\usepackage{float}
\usepackage{cite}
\usepackage{hyperref}
\usepackage{fontawesome}

\def\ba{\begin{eqnarray}}
\def\ea{\end{eqnarray}}

\def\be{\begin{equation}}
\def\ee{\end{equation}}

\begin{document}

\title{$\tt GrayHawk$: A public code for calculating the Gray Body Factors of massless fields around spherically symmetric Black Holes}

\author{ 
Marco Calz\'a$^1$\footnote{E-mail address: marco.calza@unitn.it
},\,\,\,
\\
\begin{small}
$^1$ Dipartimento di Fisica, Universit\`a di Trento, Via Sommarive 14, 38123 Povo (TN), Italy
\end{small}\\
}

\date{}

\maketitle

\abstract{
We introduce and describe $\tt GrayHawk$, a publicly available Mathematica-based tool designed for the efficient computation of gray-body factors for spherically symmetric and asymptotically flat black holes. This program provides users with a rapid and reliable means to compute gray-body factors for massless fields with spin \(s = 0, 1/2, 1, 2\) in modes specified by the angular quantum number \(l\), given a black hole metric and the associated parameter values.  

\noindent $\tt GrayHawk$ is preloaded with seven different black hole metrics, offering immediate applicability to a variety of theoretical models. Additionally, its modular structure allows users to extend its functionality easily by incorporating alternative metrics or configurations. This versatility makes $\tt GrayHawk$ a powerful and adaptable resource for researchers studying black hole physics and Hawking radiation.

\noindent The codes described in this work are publicly available at \href{https://github.com/marcocalza89/GrayHawk}{\faGithub}.

\section{Introduction}
Black holes are among the most fascinating and enigmatic objects in physics. They embody the profound interplay between gravity, quantum mechanics, and thermodynamics, offering a unique window into the fundamental laws of nature \cite{Federico:2024fyt,Maggio:2022nme,Maggio:2021ans,Calza:2024ncn,Calza:2023iqa,Calza:2022ioe,Calza:2021czr,Calza:2023rjt,Horowitz:2023xyl,Vagnozzi:2022moj,Afrin:2022ztr,Baker:2021btk,Baker:2022rkn}. Over the past few decades, black holes have risen to the forefront of theoretical and observational physics, fueled by remarkable discoveries such as the detection by LIGO-Virgo-Kagra \cite{LIGOScientific:2018jsj,LIGOScientific:2016aoc} of gravitational waves from binary mergers, the Event Horizon Telescope's groundbreaking image of a supermassive black hole \cite{EventHorizonTelescope:2019dse,EventHorizonTelescope:2022wkp}, and mounting evidence for their role in astrophysical and cosmological phenomena. These developments have highlighted black holes not only as astrophysical entities but also as testbeds for some of the deepest questions in modern physics. 

\noindent A pivotal development in black hole physics came with Hawking's discovery that black holes emit quasi-thermal radiation and therefore slowly evaporate over time \cite{Hawking:1975vcx}. This seminal work laid the foundation for a vast body of research exploring the properties of Hawking radiation and its implications for the quantum behavior of black holes. Building on Hawking's initial insights, subsequent studies by Teukolsky, Press, Page, Chandrasekhar, Detweiler, and others developed the formalism for describing perturbations of black holes for perturbing fields of various spins \cite{Newman:1961qr,Kinnersley,Kinnersley:1969zza,Pirani1964,Teukolsky:1972my,Teukolsky:1973ha,Press:1973zz,Teukolsky:1974yv,Wald 1973,Page:1976df,Page:1976ki,Page:1977um,Chandrasekhar:1975zz,Chandrasekhar:1976zz,Chandrasekhar,Chandrasekhar:1977kf}. These works yielded the Hawking's emission rates of General Relativity vacuum solutions for spin 0, 1, 2, and \(1/2\), providing the cornerstone for much of the subsequent research into black hole thermodynamics.

\noindent The importance of these studies has grown significantly in recent years as general relativity faces challenges from both cosmological and quantum perspectives. In the infrared regime, the discoveries of dark matter and dark energy have motivated modifications of general relativity, such as the inclusion of a cosmological constant and the exploration of anti-de Sitter black hole metrics with altered Hawking radiation \cite{Hemming:2000as}. In the ultraviolet regime, attempts to reconcile general relativity with quantum mechanics have led to theories such as string theory and Loop Quantum Gravity (LQG), which predict observable corrections to black hole physics. These corrections range from the presence of extra spatial dimensions \cite{Harris:2003eg,Johnson:2020tiw} to the avoidance of singularities and modifications of black hole horizons \cite{Ashtekar:2005qt,Modesto:2005zm,Boehmer:2007ket,Ashtekar:2018lag,Bodendorfer:2019jay}. Black holes, therefore, serve as natural probes for testing these corrections, translating deep quantum effects into measurable semiclassical phenomena visible from outside the event horizon.  

\noindent In this context, a wide variety of modified black hole metrics has been proposed over the last few decades \cite{Harris:2003eg,Johnson:2020tiw,Kanzi:2020cyv,Konoplya:2020jgt,Chowdhury:2020bdi,Zhang:2020qam,Konoplya:2020cbv,Rincon:2020cos,Berry:2021hos,Hayward:2005gi,Molina:2021hgx,Bardeen:1986,Simpson:2018tsi,Peltola:2008pa,Peltola:2009jm,Bianchi:2018mml,DAmbrosio:2018wgv,Sebastiani:2022wbz,Ansoldi:2008jw,Nicolini:2008aj,Torres:2022twv,Lan:2023cvz}. These modifications affect key black hole observables, such as Hawking radiation, and their measurement may serve for testing the validity of general relativity and probing the fine structure of black hole horizons. 

\noindent Within this broader context, Primordial Black Holes (PBHs) hold a special place \cite{Hawking:1971ei,Carr:1974nx,Carr:1975qj,Clesse:2017bsw,Escriva:2022duf,Sasaki:2016jop}. Formed in the early Universe from density fluctuations during periods of inflation or phase transitions, PBHs are compelling candidates for addressing some of the most pressing mysteries in cosmology. Intriguingly, PBHs in the asteroid mass range (\(10^{16}\)-\(10^{20}\, \mathrm{g}\)) are particularly significant because they could potentially account for the entire dark matter content of the Universe \cite{Katz:2018zrn,Bai:2018bej,Smyth:2019whb,Coogan:2020tuf,Ray:2021mxu,Auffinger:2022dic,Ghosh:2022okj,Miller:2021knj,Branco:2023frw,Bertrand:2023zkl,Tran:2023jci,Gorton:2024cdm,Dent:2024yje,Tamta:2024pow,Tinyakov:2024mcy,Loeb:2024tcc,Carr:2020gox,CTA}. In this case, such a PBH abundance would make it reasonable to expect that within a lifetime span a PBH would pass close enough to the Earth to enable the direct measurement of its Hawking radiation \cite{Calza:2022ioe,Calza:2023iqa,Calza:2022ljw,Calza:2023gws}. If PBHs indeed constitute dark matter, their existence would have profound implications for our understanding of cosmological structure formation, early Universe physics, and the nature of dark matter itself. 

\noindent In addition to their potential cosmological role, PBHs in this mass range exhibit unique quantum properties due to their small sizes and temperatures comparable to those of subatomic particles. Therefore, the study of their Hawking radiation is not only a window into the Standard Model (SM) and theories Beyond the Standard Model (BSM) \cite{Federico:2024fyt,Baker:2021btk,Baker:2022rkn,Calza:2024ncn,Calza:2023iqa,Calza:2021czr,Calza:2023rjt}
but also provides observable signatures that could be used to detect PBHs or constrain their properties\cite{Calza:2022ioe,Calza:2022ljw,Calza:2023gws,Calza:2024fzo,Calza:2024xdh}. 

\noindent The computation of Hawking radiation is fundamentally tied to the response of black holes to perturbations. This involves solving equations of motion for perturbing fields of various spins in the background of the black hole metric, often expressed as one-dimensional Schr\"odinger-like wave equations with short-range potentials. These potentials depend on the spin of the perturbing field and the specific properties of the black hole metric. While the formalism for Schwarzschild and Kerr black holes is well established, generalizing these results to modified gravity theories and more complex black hole metrics has become a critical area of research \cite{Berry:2021hos,Moulin:2019ekf,Arbey:2021yke,Arbey:2021jif}.  

\noindent A critical ingredient in modeling Hawking radiation are the gray body factors, which describe the partial absorption and scattering of quantum fields by the black hole's gravitational potential barrier. In practical terms, Gray-Body Factors (GBFs) modify the spectrum of Hawking radiation as it propagates from the near-horizon region to infinity, encoding information about the energy dependence of the emitted particles. Precise computation of the gray body factors is therefore essential for accurately predicting the black hole's emission spectrum, constraining observational signals, and understanding the dynamical evolution of black holes undergoing mass loss due to Hawking radiation.  

\noindent Recent years have seen a significant effort to compute gray body factors for a wide variety of black hole configurations, including Schwarzschild, Kerr, Reissner-Nordstr\"om, and more exotic metrics arising in modified gravity theories. These computations often rely on semi-analytical methods, such as the  Wentzel-Kramers-Brilloui (WKB) approximation \cite{Rincon:2020cos,Konoplya:2020cbv,Neitzke:2003mz,Singh:2024nvx,Guo:2023nkd,Konoplya:2019hlu,Gogoi:2023fow,Kanzi:2023zvo,MahdavianYekta:2019pol,Devi:2020uac}, Korteweg-de Vries (KdV) integrals \cite{Lenzi:2023inn} or employ bounding techniques to estimate the values of GBFs. While such methods provide valuable insights, they are inherently limited by the approximations involved. This limitation highlights the need for precise numerical tools capable of overcoming these challenges and providing reliable results across a broad range of configurations. 

\noindent In this paper, we present a numerical tool designed to compute gray body factors for spherically symmetric, asymptotically flat black holes. By utilizing a fully numerical approach, this tool avoids the limitations of semi-analytical methods and provides highly accurate results for black hole spectra. The tool is designed to handle a wide range of perturbing fields (spins 0, 1, 2, and \(1/2\)) and black hole metrics, making it a versatile resource for researchers studying black hole thermodynamics and Hawking radiation. In particular, the tool provides a systematic way to compute the short-range potentials and transmission coefficients associated with the Schr\"odinger-like wave equations governing black hole perturbations. 

\noindent Our goal is to equip the scientific community with a robust and user-friendly resource that can serve as a foundation for future research into the quantum properties of black holes.

\noindent This study builds upon and extends previous efforts to formulate the equations of motion for perturbing fields within generalized metrics \cite{Arbey:2021yke,Arbey:2021jif}. In this regard, it shares a conceptual alignment with the computational tool $\tt BlackHawk$. Specifically, $\tt GrayHawk$ has already demonstrated its practical relevance by contributing to the development of the forthcoming release of $\tt BlackHawk$, namely $\tt BlackHawk\;v3.0$ \cite{BHv3}.

\noindent This paper is organized as follows: In Section 2, we review the theoretical framework and background of GBFs computation and emphasize their role in BHs emission spectra. Section 3 describes the numerical methods and algorithms implemented in the tool, highlighting their accuracy, efficiency and weak points. Here we describe detailed describe the two files composing the tool. In Section 4, we validate the tool by comparing its results to existing benchmark cases present in the literature computed using a different and more involved numerical approach. In Sec.5 we provide an example of to guide the user in possible modification of the code. Finally, conclusions are drawn in Section 6. In this work we set $c=\hbar=G=1$.

\section{Theoretical formalism} \label{ThFram}
This work focuses on a particular subset of Petrov type D \cite{Petrov:2000bs} metrics. Namely, spherically-symmetric static metrics whose general form in four-dimensional Boyer-Lindquist coordinates reads
\begin{equation}\label{eq:metric}
	 ds^2=-G(r) dt^2+\frac{dr^2}{F(r)} +H(r) d\Omega^2\;,
\end{equation}
with $d\Omega^2=d\theta^2+\sin^2(\theta)d\varphi^2$ the solid angle in spherical coordinates. Furthermore, we require such line element to be asymptotically flat by imposing
\begin{equation}\label{falloffs}
	F(r)\underset{r\rightarrow+\infty}{\longrightarrow}1\;,\;\;\;\; G(r)\underset{r\rightarrow+\infty}{\longrightarrow}1\;,\;\;\;\; H(r)\underset{r\rightarrow+\infty}{\sim}r^2\,.
\end{equation}
A wide variety of BHs belong to this category, from the most known Schwarzschild and Reisner-Nordrtr\"om to the most simple regular solutions as Hayward and Bardeen BHs and even more exotic Loop Quantum Gravity (LQG) inspired Black-Bounce (BB) solutions.
\\
\noindent A further distinction in the set of spherically symmetric metrics. In fact, when
\begin{equation}\label{eq:tr_symmetric}
	G(r)=F(r)\;,\;\;\;\; H(r)=r^2\,,
\end{equation}
it is possible to dub the metric as $tr$-symmetric (time-radius symmetric). Contrary, for example in the case of BB solutions, are dubbed non-$tr$-symmetric.

\noindent Considering the Newman-Penrose (NP) fromalism and taking the null tetrad $g^{ab}=-l^an^b-n^al^b+m^a\bar{m}^b+\bar{m}^am^b$ given by
\begin{equation}
    l^a=\left(\frac{1}{G},\sqrt{\frac{F}{G}},0,0\right)\;,\;\;\;\; m^a=\left(0,0,\frac{1}{\sqrt{2H}},\frac{i}{\sqrt{2H}\sin\theta}\right)\,, \nonumber
\end{equation}
\begin{equation}
    n^a=\left(\frac{1}{2},-\frac{\sqrt{FG}}{2},0,0\right)\;,\;\;\;\;
	\displaystyle\bar{m}^a=\left(0,0,\frac{1}{\sqrt{2H}},\frac{-i}{\sqrt{2H}\sin\theta}\right)\,,\nonumber
\end{equation}
where $m$ and $\bar{m}$ are complex conjugate, satisfying $l\cdot n=-1$ and $m\cdot\bar{m}=1$ and all other scalar products vanish, one gains a grate simplification. In fact, this way, the massless field equation of different spin condense in a single master equation for the respective NP-scalars  $\Upsilon_s $

\begin{align}
    &-\dfrac{H}{G}\partial_t^2\Upsilon_s  + s\sqrt{\dfrac{F}{G}}\left( H\dfrac{G^\prime}{G} - H^\prime \right)\partial_t\Upsilon_s  + FH\partial_r^2\Upsilon_s  + \left( \dfrac{F^\prime H}{2} + (s+1/2)\dfrac{FG^\prime H}{G} + (s+1)FH^\prime \right)\partial_r \Upsilon_s  \nonumber \\
	&+\left( \dfrac{1}{\sin(\theta)}\partial_\theta (\sin(\theta)\partial_\theta) + \dfrac{2is\cot(\theta)}{\sin(\theta)}\partial_\varphi + \dfrac{1}{\sin(\theta)^2}\partial_\varphi^2 - s - s^2\cot(\theta)^2 \right)\Upsilon_s  +\Bigg[ s\dfrac{FG^{\prime\prime}H}{G} \nonumber \\ 
	&  + \dfrac{3s - 2s^2}{2}FH^{\prime\prime} - \dfrac{s}{2}\dfrac{FG^{\prime 2}H}{G^2} + \dfrac{2s^2 - s}{4}\dfrac{FH^{\prime 2}}{H} + \dfrac{s}{2}\dfrac{F^\prime G^\prime H}{G} + \dfrac{3s - 2s^2}{4} F^\prime H^\prime + \dfrac{2s^2 + 5s}{4}\dfrac{FG^\prime H^\prime}{G} \Bigg]\Upsilon_s  = 0 \label{eq:Teukolsky_master}
\end{align}
Under the Ansatz
\begin{equation}\label{eq:anzatz}
	\Upsilon_s (t,r,\theta,\varphi)=\Phi_s(r)S^s_{l,m}(\theta,\varphi)e^{-i\omega t}\,,
\end{equation}
Eq.(\ref{eq:Teukolsky_master}) separates in the spin weighted spherical harmonics equation satisfying \cite{Chandrasekhar1990,Kalnins:1992,Fackerell&Crossman1977,Suffern1983,Seidel:1988ue,Berti:2005gp}
\begin{equation}
	\left(\frac{1}{\sin\theta}\partial_\theta(\sin\theta\,\partial_\theta)+\csc^2\theta\,\partial_\varphi^2+\frac{2is\cot\theta}{\sin\theta}\partial_\varphi+s-s^2\cot^2\theta+\lambda_l^s\right)S_{l,m}^s=0\,,
\end{equation}
with $\lambda_l^s= l(l+1)-s(s+1)$ separation constant, and the radial part satisfying
\begin{equation}\label{eq:teukolsky_general}
	A_s\big(B_s\Phi'_s\big)'+\left(\frac{H}{G}\omega^2+i\omega s\sqrt{\frac{F}{G}}\left(H'-H \frac{G'}{G}\right)+C_s\right)\Phi_s=0\,,
\end{equation}
where
\begin{eqnarray}
A_s= \sqrt{\frac{F}{G}} \frac{1}{(G H)^s}\,,
\label{eq:as}
\end{eqnarray}
\begin{eqnarray}
B_s=\sqrt{F G } (G H)^s H\,,
\end{eqnarray}
\begin{align}
C_s &= s \frac{F H G''}{G} + \frac{s}{2} \left ( \frac{H F' G' }{G} - \frac{F H G'^2}{G^2} \right ) \nonumber \\ 
&+ \frac{s(3-2s)}{4} \left( 2 F H'' + F' H'  \right) +\frac{s(2s-1)}{4} \frac{F H'^2}{H}\nonumber \\ 
&+\frac{s(2s+5)}{4}\frac{F G' H'}{G}-\lambda^s_l-2s\,.
\end{align}
Adopting tortoise coordinate
\begin{equation}\label{tort}
    \frac{dr^*}{dr} = \frac{1}{\sqrt{F G}}
\end{equation}
and with further manipulations the radial equation may be written in Schr\"odinger-like form reading
\begin{equation} \label{SchEq}
    \partial_{r^*}Z_s + \Bigl(\omega^2-V_s(r(r^*))\Bigr)Z_s =0
\end{equation}
where
\begin{subequations}\label{eq:potentials}
\begin{align}
	&V_0=\nu^0_l\frac{G}{H}+\frac{\partial_{r^*}^2\sqrt{H}}{\sqrt{H}}\,,\\
	&V_1=\nu^1_l\frac{G}{H}\,,\\
	&V_2=\nu^2_l\frac{G}{H}+\frac{(\partial_{r^*}H)^2}{2H^2}-\frac{\partial_{r^*}^2\sqrt{H}}{\sqrt{H}}\,,\\
	&V_{1/2}=\nu^{1/2}_l\frac{G}{H} \pm \sqrt{\nu^{1/2}_l}\,\partial_{r^*}\left( \sqrt{\frac{G}{H}} \right)\,,
\end{align}
\end{subequations}
and $\nu^s_l=l(l+1)-s(s-1)$.\\
\noindent Given the choices (\ref{falloffs}) and the requirement that the metric describes a BH (Namely, the metric has an event horizon and if asymptotically flat), $V_s$ vanishes at the horizon ($r^* \rightarrow -\infty$) and at infinity ($r^* \rightarrow +\infty$). This way the asymptotic solutions read
\begin{subequations}
\begin{align}
	&Z_s(r^*\rightarrow -\infty) = \mathfrak{a} \;e^{i \omega r^*}+\mathfrak{b} \;e^{-i \omega r^*} \label{asympt-} \,,\\
    &Z_s(r^*\rightarrow +\infty) = a \;e^{i \omega r^*}+b \;e^{-i \omega r^*} \label{asympt+}\,.
\end{align}
\end{subequations}
On the event horizon, we invoke purely in-going boundary conditions and normalize the wave function to unity. Namely, $\mathfrak{a}=0$ and $\mathfrak{b}=1$.

\noindent Under such choices the scattering problem  transmission and reflection coefficients read
\begin{equation}
    R=\frac{1}{|a|^2}\;\;\;T=\frac{1}{|b|^2}\;.
\end{equation}
Those coefficients are functions of the energy and depend on the field spin and mode. The transmission coefficient, measuring the variation from a BH purely black-body emission, is often called gray-body factor and denoted with $\Gamma_s^l(\omega)$.\\
Chosen a specific field $s$ and mode $l$ and given in-going boundary condition and normalization, the function $Z_s$ is known at the horizon and with the aid of Eq.(\ref{SchEq}) it is possible to integrate out the solution in the region far away form the horizon obtaining $\Gamma_s^l(\omega)$.\\
The number of particles emitted per unit time and energy emitted by Hawking radiation for a given particle species $i$ with spin $s$, as a result of Hawking evaporation ~\cite{Hawking:1975vcx,Page:1976df,Page:1976ki,Page:1977um}, is given by :~\footnote{This expression implicitly assumes that the particles emitted by the BH are not coupled to the regularizing parameter $\ell$, an assumption which is reasonable.}
\begin{eqnarray}\label{prim}
\frac{d^2N_i}{dtdE_i}=\frac{1}{2\pi}\sum_{l,m}\frac{n_i\Gamma^s_{l,m}(\omega)}{ e^{\omega/T}\pm 1}\,,
\label{eq:d2ndtdei}
\end{eqnarray}
with the plus (minus) sign in the denominator is associated to fermions (bosons), and where $n_i$ is the number of degrees of freedom of the particle in question, $\omega=E_i$ is the mode frequency (in natural units). We implicitly set $k_B=1$.
Finally, the temperature is related to the surface gravity computed at the horizon and reads
\begin{eqnarray}
T=\sqrt{\frac{F(r)}{G(r)}}\frac{G'(r)}{4\pi}\vert_{r_H}\,,
\label{eq:temperature}
\end{eqnarray}

\subsection{Implemented metrics}
We report here the line element proposed in the code. It is important to emphasize that alternative line elements can also be considered, provided they satisfy the requirement of describing an asymptotically flat black hole.
\subsubsection{Schwarzschild}
The Schwarzschild spacetime is the first discovered and simplest exact solution to Einstein's field equations, representing a static, spherically symmetric black hole with no charge or rotation. It describes the gravitational field surrounding a non-rotating, uncharged massive object in General Relativity and is given by the line element:  
\[
ds^2 = -\left(1 - \frac{2M}{r}\right) dt^2 + \frac{dr^2}{1 - \frac{2M}{r}} + r^2 d\Omega^2\,,
\]  
Here, \(M\) represents the mass of the black hole, and the parameter \(r\) denotes the radial coordinate, which corresponds to the circumferential radius.  
The Schwarzschild black hole has a single event horizon located at \(r_H = 2M\). This is the radius at which the escape velocity equals the speed of light, effectively trapping all matter and radiation within. 
At \(r = 0\), the spacetime curvature becomes infinite, leading to a singularity. This singularity is hidden within the event horizon and cannot be observed from the outside. 
  
\noindent The Schwarzschild solution is a cornerstone of black hole physics, serving as the prototype for understanding more complex black hole spacetimes. Its simplicity makes it a critical testbed for exploring general relativistic effects, such as gravitational time dilation, light bending, and redshift.  

\noindent While the Schwarzschild black hole is an idealized solution, it provides a foundational model for studying the fundamental properties of black holes and their interaction. On top of this the Schwarzschild BH is the limit case of more complex BHs when all parameters other the M are vanishing.

\subsubsection{Reissner-Nordstr\"om}
The Reissner-Nordstr\"om (RN) spacetime describes a static, spherically symmetric black hole that incorporates both mass \(M\) and electric charge \(Q\). It is a solution to the Einstein-Maxwell equations in general relativity and generalizes the Schwarzschild solution by including the effects of the electromagnetic field. The line element for the RN metric is expressed as:  
\[
ds^2 = -\left(1 - \frac{2M}{r} + \frac{Q^2}{r^2}\right) dt^2 + \frac{dr^2}{1 - \frac{2M}{r} + \frac{Q^2}{r^2}} + r^2 d\Omega^2\,,
\]

\noindent The RN spacetime exhibits a rich structure, with the possibility of multiple horizons. Specifically, it can feature (a) two horizons when \(Q^2 < M^2\), the spacetime has an outer event horizon at \(r_+\) and an inner Cauchy horizon at \(r_-\), given by \(r_\pm = M \pm \sqrt{M^2 - Q^2}\), (b) one degenerate horizon when \(Q^2 = M^2\), the spacetime represents an extremal black hole, where the two horizons coincide at \(r = M\), (c) no horizons when \(Q^2 > M^2\), the solution corresponds to a naked singularity, violating the cosmic censorship conjecture.  

\noindent The RN metric reduces to the Schwarzschild solution when \(Q = 0\) and becomes flat Minkowski spacetime as both \(M\) and \(Q\) vanish.  

\noindent Physically, the RN spacetime describes a black hole with an electric field, and its singularity at \(r = 0\) remains unresolved within classical general relativity. However, it provides valuable insights into the interplay between gravity and electromagnetism and serves as a starting point for exploring charged black holes in modified theories of gravity or quantum gravity frameworks.  
\subsubsection{Hayward}
The Hayward regular BH ~\cite{Hayward:2005gi} is one of the most well-known RBHs. Its line element is given by: 
\[
ds^2=\left(1 - \frac{2Mr^2}{r^3 + 2M\ell^2}\right) dt^2 + \frac{dr^2}{1 - \frac{2Mr^2}{r^3 + 2M\ell^2}} + r^2 d\Omega^2\,,
\label{eq:frhayward}
\]

\noindent Like the following Bardeen RBH, the Hayward RBH features a de Sitter (dS) core that replaces the central singularity. In fact, the introduction of a dS core, characterized by a positive cosmological constant \(\Lambda = 3/\ell^2\), was the original motivation for the Hayward BH, which was initially proposed on purely phenomenological grounds. Despite its phenomenological origins, theoretical frameworks have been explored to explain the Hayward BH. These include corrections to the equation of state of matter at high densities~\cite{Sakharov:1966,Gliner:1966}, finite density and curvature models~\cite{Markov:1982,Markov:1987,Brandenberger:1992sy}, non-linear electrodynamics theories~\cite{Kumar:2020bqf,Kruglov:2021yya}, and quantum gravity corrections~\cite{Addazi:2021xuf,AlvesBatista:2023wqm}. We treat the Hayward RBH here as a model-agnostic phenomenological toy model of a singularity-free spacetime.
\subsubsection{Bardeen}
Another well-known regular spacetime, and among the first proposed~\cite{Bardeen:1986}, is the Bardeen BH
\[
ds^2=\left(1 - \frac{2Mr^2}{(r^2 + \ell^2)^{3/2}}\right) dt^2 + \frac{dr^2}{1 - \frac{2Mr^2}{(r^2 + \ell^2)^{3/2}}}  + r^2 d\Omega^2\,,
\label{eq:frbardeen}
\]  
where \(M\) represents the BH mass, which can be consistently identified with various definitions of mass (such as Komar, ADM, Misner-Sharp-Hernandez, or Brown-York mass).

\noindent As for the Hayward BH, the regularization parameter \(\ell\) must satisfy \(\ell \leq \sqrt{16/27} M \sim 0.77 M\) to ensure that the metric describes a BH rather than a horizonless object. For \(\ell \to 0\), the metric function reduces to that of the Schwarzschild BH. 

\noindent The Bardeen BH notably features a de Sitter (dS) core that replaces the central singularity of the Schwarzschild BH. This is evident from the metric behavior in the \(r \to 0\) limit, where \(F(r)=G(r) \propto r^2\), consistent with an asymptotically dS spacetime. Initially introduced on phenomenological grounds, the Bardeen RBH is now known to arise from a magnetic monopole source~\cite{Ayon-Beato:2000mjt}, potentially within certain non-linear electrodynamics theories~\cite{Ayon-Beato:2004ywd}. It may also originate from quantum corrections to the uncertainty principle~\cite{Maluf:2018ksj}. As with the Hayward RBH, rgardless of its origin, and consistent with the approach taken for other spacetimes, this solution is treated as a model-agnostic phenomenological toy model.

\subsubsection{Simpson-Visser}
The Simpson-Visser (SV) metric is a one-parameter extension of the Schwarzschild spacetime and is among the most well-known black-bounce metrics. According to Simpson and Visser, this metric represents the minimal violence to the standard Schwarzschild solution necessary to ensure regularity~\cite{Simpson:2018tsi}. It analytically interpolates between black holes (BHs) and traversable wormholes (WHs), depending on the value of the regularizing parameter. 
The SV line element is given by ~\footnote{For all spacetimes discussed, the parameter \(M\) consistently corresponds to the mass (e.g., Komar, ADM, Misner-Sharp-Hernandez, or Brown-York mass).}  :  
\[
ds^2 = - \left(1 - \frac{2M}{\sqrt{r^2 + \ell^2}} \right) dt^2 + \frac{dr^2}{1 - \frac{2M}{\sqrt{r^2 + \ell^2}}} + (r^2 + \ell^2) d\Omega^2\,.
\label{eq:metricsimpsonvisser}
\]  
The SV spacetime exhibits a rich phenomenology, interpolating between different configurations based on the ratio \(\ell/M\). It reduces to the Schwarzschild BH for \(\ell = 0\), describes a regular BH with a one-way space-like throat for \(0 < \ell/M < 2\), transitions to a one-way wormhole with an extremal null throat at \(\ell/M = 2\), and becomes a traversable WH with a two-way time-like throat for \(\ell/M > 2\).~\footnote{These cases arise from analyzing: (a) the existence of horizons, determined by the zeros of $F(r)$, which yield \(r_H = \sqrt{4M^2 - \ell^2}\) and require \(\ell < 2M\) for a BH, and (b) the sign of the coordinate speed of light, \(\vert dr/dt \vert = 1 - 2M/\sqrt{r^2 + \ell^2}\), which determines whether the wormhole for \(\ell \geq 2M\) is traversable (\(dr/dt \neq 0\) for all \(r\)) or not (\(dr/dt \to 0\) as \(r \to 0\)). For a detailed explanation, refer to the discussion above Eq.~(2.6) in Ref.~\cite{Simpson:2018tsi}.}  
This metric has inspired numerous subsequent studies (e.g., Refs.~\cite{Tsukamoto:2020bjm,Mazza:2021rgq,Shaikh:2021yux,Islam:2021ful,Guerrero:2021ues,Bambhaniya:2021ugr,Yang:2022xxh,Riaz:2022rlx,Arora:2023ltv,Jha:2023wzo,Jha:2023nkh}). While originally proposed on phenomenological grounds, it may arise as a solution of GR coupled to non-linear electrodynamics with a minimally coupled phantom scalar field~\cite{Bronnikov:2021uta}.  

\subsubsection{Peltola-Kunstatter}
The Peltola-Kunstatter (PK) spacetime is a metric inspired by LQG, derived by applying effective polymerization techniques to four-dimensional Schwarzschild black holes. While LQG has been suggested as a framework capable of resolving the singularities inherent in GR, the complexity of solving the full system has spurred the development of semiclassical polymer quantization techniques. These methods, which are unitarily inequivalent to Schr\"odinger quantization, preserve the essential feature of spacetime discreteness. The PK spacetime results from polymerizing the area while leaving the conformal mode unmodified. This leads to a regular spacetime where the singularity is replaced by a complete and nonsingular bounce, with the metric contracting to a minimum radius before expanding into a Kantowski-Sachs geometry~\cite{Peltola:2008pa,Peltola:2009jm}.  
The line element takes the form:  
\[
ds^2 = - \left(\frac{r - 2M}{\sqrt{r^2 + \ell^2}} \right) dt^2 + \frac{dr^2}{\frac{r - 2M}{\sqrt{r^2 + \ell^2}}} + (r^2 + \ell^2) d\Omega^2\,.
\label{eq:metricpeltolakunstatter}
\]  
In what follows, we treat the PK spacetime as an example of a regular metric derived from first-principles quantum gravity considerations, distinguishing it from the phenomenological metrics discussed earlier.  

\subsubsection{D' Ambrosio-Rovelli}
The D'Ambrosio-Rovelli (DR) spacetime, though not originally designed to address singularity avoidance, is also rooted in LQG considerations. This spacetime naturally extends the Schwarzschild solution by smoothly traversing the \(r=0\) singularity into the interior of a white hole. It can be viewed as the \(\hbar \to 0\) limit of an effective quantum gravity metric~\cite{Bianchi:2018mml,DAmbrosio:2018wgv}. This black hole-to-white hole tunneling mechanism has been proposed as a potential resolution to the information paradox. For our purposes, the DR metric is notable for being regular, as the curvature of its effective metric remains bounded at the Planck scale.  

\noindent The ansatz for the effective metric proposed by D'Ambrosio and Rovelli is structurally similar to the Simpson-Visser (SV) metric but differs in the form of the $F(r)$ function. The line element for the DR spacetime is:  
\[
ds^2 = - \left(1 - \frac{2M}{\sqrt{r^2 + \ell^2}} \right) dt^2 + \frac{dr^2}{\left(1 - \frac{2M}{\sqrt{r^2 + \ell^2}}\right) \left(1 + \frac{\ell}{\sqrt{r^2 + \ell^2}}\right)} + (r^2 + \ell^2) d\Omega^2\,. \nonumber
\]  

Similarly to the Peltola-Kunstatter (PK) spacetime, we consider the DR spacetime to be a well-motivated example of a quantum gravity-inspired metric. Additionally, the assumption of primordial DR black holes implies the existence of long-lived primordial DR white holes because of quantum transitions occurring near the would-be singularity. This could lead to intriguing phenomenology, although exploring these implications lies beyond the scope of this work.

\section{Numerical methodology and code description}

The code is organized into two files. The primary file, $\tt GrayHawk.nb$, is designed to compute the gray-body factor as a function of energy, expressed in units of the black hole mass \( M \). The secondary file, $\tt CalibratorGH.nb$, serves as a supplementary tool to assist with calibration and parameter selection.

\subsection{$\tt GrayHawk.nb$}
Here, we describe the primary tool used to compute GBFs. The program consists of a single, self-contained piece of code that can be executed with a single command. Upon execution, the tool generates several outputs, including: a plot of the geometric potential, a table displaying the computed GBF values, and a corresponding GBF plot.  

\noindent To facilitate organization and usability, the code is structured into five main sections, each divided into specific subsections. The details of these sections are described below.
\begin{itemize}
    \item \textbf{Parameters} In this section, the code is initialized with the explicit values of the parameters defining both the field and the metric, as well as the energy values at which the GBFs are to be evaluated. It is important to note that the mass parameter \( M \) must be set to one, as the code is designed to operate in units normalized to the black hole mass.
    \begin{itemize}
        \item \textit{\textbf{BH parameters}}
        This subsection is dedicated to specifying the parameters that define the metric. If the user wishes to implement a metric other than the pre-compiled options, any additional parameter(s) associated with the new metric must be specified in this section.
        \item \textit{\textbf{Field parameters}} In this section, users can specify the spin of the field for which the GBF is to be calculated, as well as the field mode in the spherical harmonic decomposition.
        \item \textit{\textbf{Energy table}} It is defined here the table of energies at which the GBF have to be probed.
    \end{itemize}
    \item \textbf{The metric} Given the line element specified in (\ref{eq:metric}), the metric functions $F(r)$, $G(r)$, and $H(r)$ are defined in the respective subsections as follows:
    \begin{itemize}
        \item \textit{\textbf{Definition of $F(r)$}}
        \item \textit{\textbf{Definition of $G(r)$}}
        \item \textit{\textbf{Definition of $H(r)$}}
    \end{itemize}
    A set of pre-compiled metrics is provided for user convenience. To utilize these pre-compiled metrics, the user simply needs to uncomment the desired metric consistently across the three metric functions (\( F(r) \), \( G(r) \), and \( H(r) \)). If the user wishes to implement a custom line element that is not pre-compiled, it can be introduced in this section. However, it is important to note that altering the line element does not guarantee accurate results. Any newly introduced line element must represent an asymptotically flat black hole, providing a well-defined event horizon, and adhere to the conditions specified in (\ref{falloffs}).
    \item \textbf{Numerical inversion of tortoise coordinates}
    To explicitly express (\ref{SchEq}), it is necessary to determine \( V_s(r(r^*)) \) and, consequently, \( r(r^*) \). In this section, we implement a numerical procedure that generates a table of \( r \) values as a function of \( r^* \). This table is subsequently interpolated to obtain a continuous representation of \( r(r^*) \).
    \begin{itemize}
        \item \textbf{\textit{Definition of the horizon radius}} The horizon radius $r_H$ is obtained as the largest root of $F(r)$.
        \item \textbf{\textit{Definition of the tortoise coordinate}} The tortoise coordinate \( r^* \) is obtained by integrating (\ref{tort}). The code utilizes Mathematica's built-in $\tt{Integrate[] }$ function to compute the indefinite integral of the given expression. However, modifying the metric may lead to complications with this procedure. For instance, when considering a Culetu-Ghosh-Simpson-Visser black hole \cite{Ghosh:2014pba,Culetu:2014lca,Simpson:2019mud}, the integral $\tt{Integrate[1-\frac{2M}{r}\exp \left ( -\frac{\ell}{r} \right )]}$ does not yield an explicit analytic result. In such cases, a purely numerical integration approach may need to be employed.
        \item \textbf{\textit{Numerical inversion}} The sampling of \( r^*(r) \) is divided into two distinct regimes: one near the horizon and another in the far-away region. It is crucial that these two regimes overlap in the central region, ensuring continuity when combined. The sampling criteria are flexible and depend on the user-defined parameters specified in the first section. Users are advised to verify the efficiency of the sampling using the $\tt CalibratorGH.nb$ tool to ensure compatibility with their chosen energy and parameter values, as discussed further in Sec. (\ref{Calib}). The two sampling regimes are subsequently merged and interpolated within the boundaries of the sampled region. This subsection is organized into the following parts:  
        \begin{itemize}
            \item \textit{Near-horizon sampling}
            \item \textit{Far-horizon sampling}
            \item \textit{Inverted tortoise coordinates}
        \end{itemize}
    \end{itemize}
    \item \textbf{The Geometrical potential}
    In this section, we utilize the definitions provided in Eq. (\ref{eq:potentials}) along with the results obtained from the previous section to explicitly compute \( V_s(r(r^*)) \). Additionally, this section includes a plot of the geometric potential, which serves as a diagnostic tool to verify whether the potential vanishes at the boundaries of the region of interest. If the potential does not approach zero at the boundaries, adjustments to the sampling described in the previous section may be required. The following subsections are part of this section:  
    \begin{itemize}
        \item \textbf{\textit{Auxiliary functions}}
        \item \textbf{\textit{Geometric potential definition}}
        \begin{itemize}
            \item \textit{Geometric potential plot}
        \end{itemize}
    \end{itemize}
    \item \textbf{Solving for the GBF} This section implements the core part of the calculation, focusing on the table of energies at which the GBF is to be evaluated. A $\tt for$ loop is employed to solve the scattering problem for each energy listed in the table. The procedure for solving the scattering problem follows the framework outlined in Sec. (\ref{ThFram}): the radial equation (\ref{SchEq}) is solved by imposing purely ingoing boundary conditions and normalizing the wave function at the event horizon, a region in the asymptotic far-away domain is defined, where the computed solution is sampled and fitted using the asymptotic expression (\ref{asympt+}), the GBF is computed as \( \Gamma = 1/|b|^2 \), where \( b \) is obtained from the far-field fit.
    
    \noindent At the conclusion of the $\tt for$ loop, the calculated values of \( \Gamma \) are compiled into a displayed table and subsequently plotted for visualization.
\end{itemize}

\noindent As default setting the code parameters are all set to 0 with the exception of the BH mass, which must always be set to one, and the line element considered is the Schwarzschild one. The table of energies to be probed is made up of 100 equidistance values between 0 and 1 (in units such that $c=\hbar=G=M=1$).
The near-horizon sampling consists of 998 logarithmically distributed points between $r_H (1+10^{-11})$ and $r_H (1+0.1)$, while the far-horizon sampling consists of 4998 equidistance points between $r_H (1+0.2)$ and $r_H (1+4.999)$.
Finally the far-horizon portion of spacetime where to sample the function $Z_s$ is taken between the maximum value of the sampled tortoise coordinate $r^*(r=r_H (1+4.999))$ and $3/4$ of this distance.

\noindent Given such settings, the code provides all the output in roughly less than 7 to 10 seconds on modern laptops.

\noindent Any change in the setting results in different computation times, for example, if one considers a D'Ambrosio-Rovelli BH, a spin-2 field and a mode $l=3$ leaving unchanged the other settings, the computational time reach roughly 20 seconds.

\subsection{$\tt CalibratorGH.nb$}
\label{Calib}
This supplementary file is provided to assist users in accurately defining and sizing the near-horizon and far-away sampling regions for the tortoise coordinate for reconstructing the potential (\ref{eq:potentials}), as well as selecting an appropriate sampling frequency and interval in the far-away region to obtain $Z_s$.  

\noindent The provided code operates similarly to the main code but, instead of iterating over the full range of energies using a $\tt for$ loop, it focuses only on the minimal and maximal energy values, \(\omega_{min}\) and \(\omega_{max}\). As with the main code, the first output is a plot of \(V_s(r(r^*))\), which allows users to verify whether the chosen minimum and maximum values of \(r^*\) in the sampling region are sufficient to ensure that the potential vanishes in both asymptotic regions. Subsequently, the code generates plots of \(Z_s(r^*)|_{\omega_{min}}\) and \(Z_s(r^*)|_{\omega_{max}}\), alongside the corresponding values \(\Gamma_s^l(\omega_{min})\) and \(\Gamma_s^l(\omega_{max})\). These outputs enable an analysis of two key scenarios to ensure that the sampling parameters are appropriately chosen for the specified energy range.  

\paragraph{Case 1: Sampling at \(\omega_{min}\) }
Inadequate sampling for \(\omega_{min}\) may result in \(Z_s(r^*)|_{\omega_{min}}\) failing to exhibit oscillatory wave behavior in the far-away region (large values of \(r^*\)), leading to an overestimation of \(\Gamma_s^l(\omega_{min})\). If this occurs, the sampling region in the far-away zone should be extended to include spacetime regions further from the horizon.

\paragraph{Case 2: Sampling at \(\omega_{max}\)  }
For \(\omega_{max}\), poor sampling may manifest as oscillatory but growing behavior in the far-away region. This indicates an inadequate choice of the near-horizon sampling distance. Specifically, if the near-horizon sampling distance is not sufficiently small, the starting point for outward integration is suboptimal, resulting in an overestimation of \(\Gamma_s^l(\omega_{max})\), which should always remain below \(1\). This issue can be resolved by selecting smaller near-horizon sampling distances.
\\
\\
\noindent In both scenarios, the parameter defined as $x_{far}$ in the code should be chosen carefully to ensure that the spacetime region between \(x_{max\infty} - x_{far}\) and \(x_{max\infty}\) lies fully within the far-away region. Additionally, the sampling frequency must satisfy the Nyquist-Shannon sampling theorem to avoid aliasing.  

\noindent For energy ranges that span several orders of magnitude, it may be computationally efficient to divide the calculation into multiple sections, each with tailored sampling parameters, to reduce the overall computational cost while maintaining accuracy.

\section{Validation and Comparison}

In this section, we present a comparison of our results with those provided in \cite{Calza:2024fzo,Calza:2024xdh}. Rather than highlighting discrepancies with respect to semi-analytical approximations, we validate our findings by benchmarking them against the most accurate spectra currently available in the literature. Specifically, \cite{Calza:2024fzo,Calza:2024xdh} report the spectra for Hayward, Bardeen, SV, KP, DR black holes, where the GBFs are computed in units of the horizon radius \(r_H\). These computations employ the Frobenius method to solve the scattering problem after rewriting the radial equation in a rescaled radial coordinate. While this approach is highly refined, it is also computationally and mathematically more intensive.  

\noindent To facilitate the comparison, we first converted the regularization parameter values provided in \cite{Calza:2024fzo,Calza:2024xdh} (expressed in units of \(r_H\)) into equivalent values in units of the black hole mass \(M\), as summarized below:  

\begin{itemize}
    \item Hayward: $\ell=0.15 r_h \Rightarrow \ell=0.293 M \;\ell=0.3 r_h \Rightarrow \ell=0.546 M \;\ell=0.45 r_h \Rightarrow \ell=0.718 M$
    \item Bardeen: \,\,$\ell=0.15 r_h \Rightarrow \ell=0.290 M \;\ell=0.3r_h \Rightarrow \ell=0.527 M \;\ell=0.45r_h \Rightarrow \ell=0.683 M$
    \item SV: \;\;\;\;\;\;\;\;\;\,$\ell=0.3 r_h \Rightarrow \ell=0.575 M \;\ell=0.6 r_h \Rightarrow \ell=1.029 M \;\ell=0.9 r_h \Rightarrow \ell=1.338 M$
    \item PK:\;\;\;\;\;\;\;\;\;\;$\ell=0.3 r_h \Rightarrow \ell=0.6 M \;\ell=0.6 r_h \Rightarrow \ell=1.2 M \;\ell=0.9 r_h \Rightarrow \ell=1.8 M$
    \item DR:\;\;\;\;\;\;\;\;\;\;$\ell=0.3r_h \Rightarrow \ell=0.575M \;\ell=0.6r_h \Rightarrow \ell=1.029 M \;\ell=0.9r_h \Rightarrow \ell=1.338M$
\end{itemize} 

\noindent Fig.(1) and Fig.(2) illustrate a direct comparison between the photon spectra obtained using $\tt GrayHawk$ and the results published in \cite{Calza:2024fzo,Calza:2024xdh}. In both cases, the spectra are calculated according to Eq.(\ref{prim}) and include contributions up to \(l=4\). Due to a computational complication the D'Ambrosio-Rovelli BH spectra in \cite{Calza:2024xdh} are  imprecise. This is not compromising the reasoning drown in that paper, nonetheless forced us to re-compute the spectra with the method described there. Therefore, in the rightmost plot of Fig.(2), we compare the results obtained with $\tt GrayHawk$ and the one obtained with the method of \cite{Calza:2024xdh} once adjusted the computational complication. The superimposed plots reveal an excellent agreement: the spectra overlap to such an extent that they are visually indistinguishable. 

\begin{figure}[]
\centering
\begin{minipage}{.49\textwidth}
  \centering
  \includegraphics[width=0.8\linewidth]{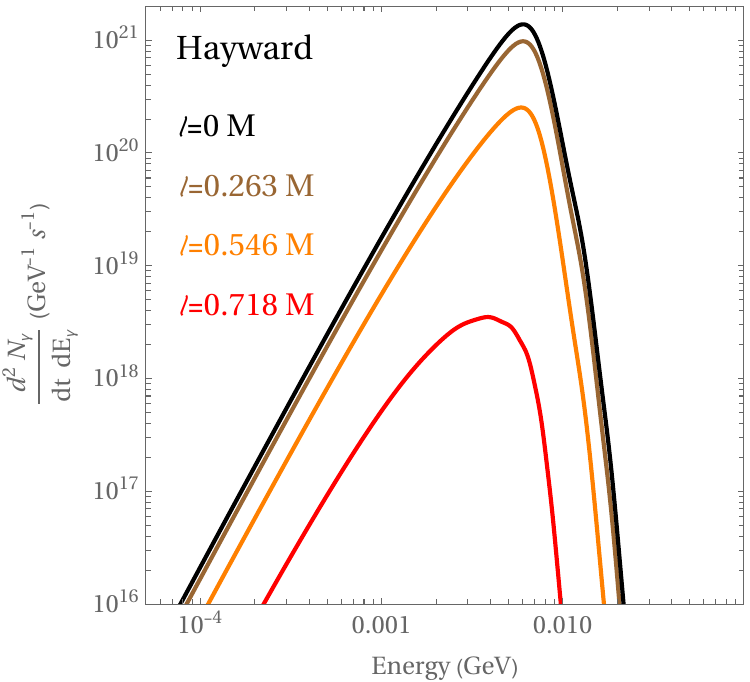}
\end{minipage}%
\begin{minipage}{.49\textwidth}
  \centering
  \includegraphics[width=0.8\linewidth]{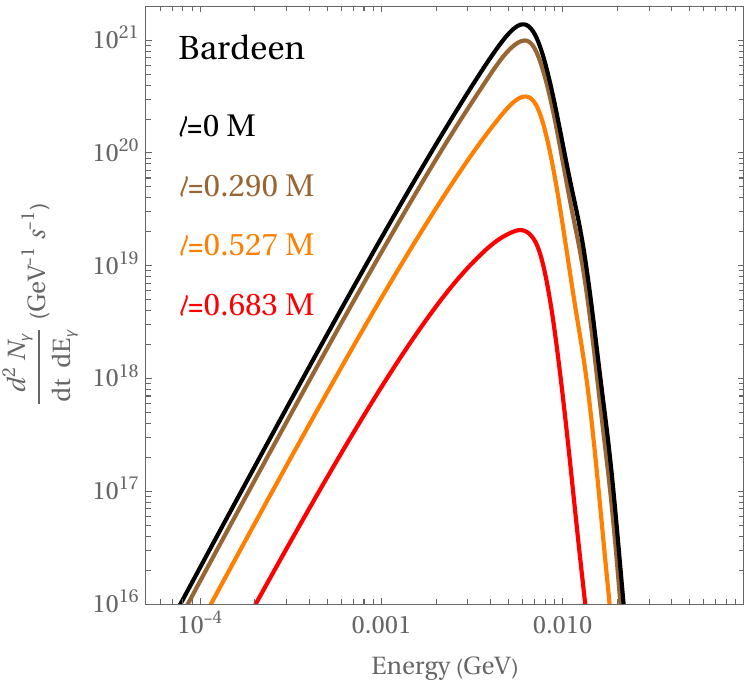}
\end{minipage}\caption{Comparison of the spectra reported in \cite{Calza:2024fzo} and the one computed using $\tt GrayHawk$ for the Hayward and Bardeen BHs}
\end{figure}

\noindent To quantify the agreement, we evaluated the discrepancies by focusing on the position of the spectral peak. Specifically, we calculate the residual R as the absolute value of the difference, weighted over the average.
Fig.(3) shows the residual of the intensity at a given energy computed for a Hayward BH of $M=10^{16}g$ and $\ell=0.3 r_H=0.546 M$.
\begin{figure}[]
\centering
\begin{minipage}{.32\textwidth}
  \centering
  \includegraphics[width=0.95\linewidth]{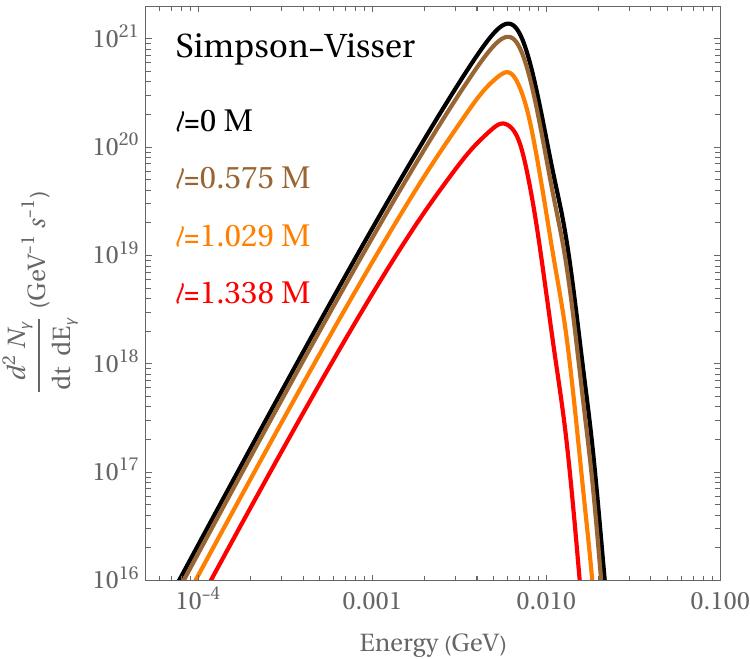}
\end{minipage}%
\begin{minipage}{.32\textwidth}
  \centering
  \includegraphics[width=0.95\linewidth]{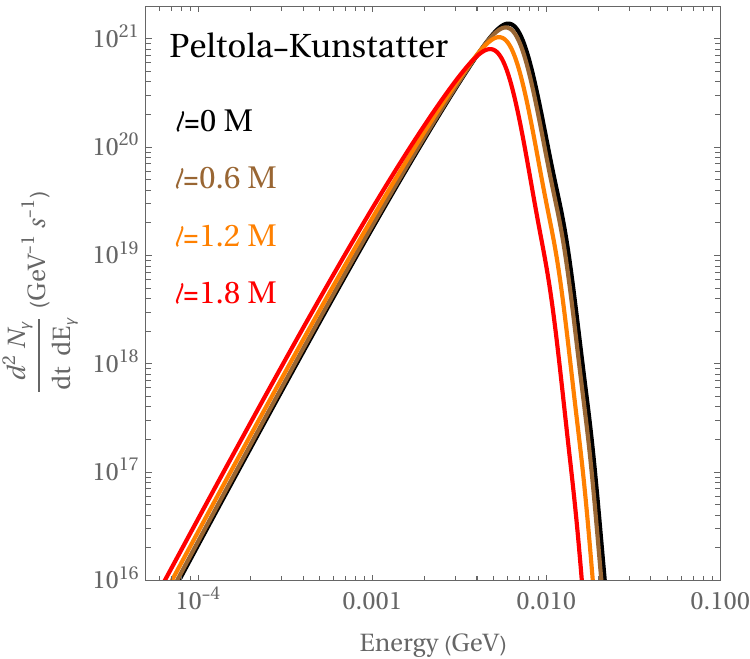}
\end{minipage}
\begin{minipage}{.32\textwidth}
  \centering
  \includegraphics[width=0.95\linewidth]{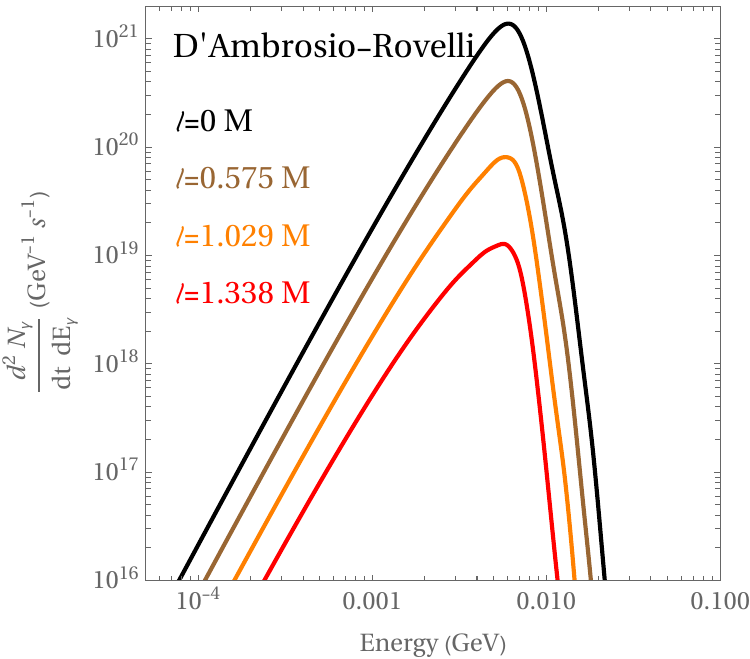}
\end{minipage}\caption{Comparison of the spectra reported in \cite{Calza:2024xdh} and the one computed using $\tt GrayHawk$ for the SV,PK, and DR BHs}
\end{figure}
Fig.(3) highlights that the two spectra diverge one with respect to the other in less then one part in 1000 throughout a window large more than two orders of magnitude in energies and centered in the most dominant part of the spectrum, the peak.
One may also consider the residual at the peak position for both energy and intensity. In general, the energy discrepancies at the peak are less than \(1\%\), while the intensity discrepancies are below \(1\text{\textperthousand}\).

\begin{figure}[]
\centering
\includegraphics[width=0.8\linewidth]{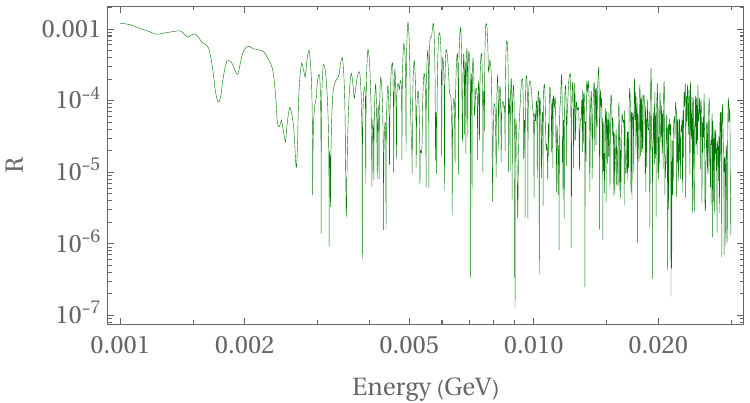}
\caption{Residual of the spectrum of a Hayward BH of $M=10^{16}g$ and $\ell=0.3r_H=0.546M$ computed with $\tt GrayHawk$ and the one from \cite{Calza:2024fzo}}
\end{figure}
\section{How to modify}
This section is intended to be a very brief guide on how to work hands on the code in order to modify it to consider different metrics. Instead of providing elaborated recipes we make use of  an example to show how this process should be done.

\noindent Let us assume that the user is interested in obtaining the leading emission mode for photons ($l=s=1$) GBFs of a Frolov regular BH \cite{Frolov:2016pav}. This type of BH is a generalization of the Hayward BH with an additional charge parameter, $q$, inspired by quantum-gravity arguments. Let us also assume that the user is interested in the following values $\ell=0.3$ and $q=0.2$ (at $c=\hbar=G=M=1$).
The Frolov BH is a $tr$-symmetric BH characterized by the line element
\[
ds^2=\left(1 - \frac{(2M-q^2)r^2}{r^3 + 2M\ell^2}\right) dt^2 + \frac{dr^2}{1 - \frac{(2M-q^2)r^2}{r^3 + 2M\ell^2}} + r^2 d\Omega^2\,.
\label{eq:frhayward}
\]
To further schematize the procedure, let us break it down into a list of actions.
The modifications with respect the original files are reported and underlined in orange in the file $\tt exampleGH.nb$ which can also be found in the repository containing the main code and the calibrator.
We stat by taking into account the calibrating code, to ensure whether or not the scattering problem is solved with a good choice of sampling, given the different metric and parameters.
\begin{itemize}
    \item Since the metric under exam is defined by an additional parameter, the first step is to introduce the parameter in the list of the BH parameters. This can simply be done adding a line at the section "BH parameters", in which $q$ is defined and initialized to the desired value $q=0.2$.
    \item Second step is to initialize the other parameters in to the desired values. For doing so we set $\ell=0.3$, $l=s=1$.
    \item The table of probed energy is left untouched in this example.
    \item The metric needs to be introduced in the code and it must be said to the code to consider the Frolov line element. We do so by commenting the line element previously considered in the definition of the functions $F(r)$, $G(r)$ (in this case the Schwarzschild one) and introducing the new metric functions for the Frolov matric by adding two new lines in the code.
    \item Since the Frolov metric is $tr$-symmetric the function $H(r)$ is the same as in the Schwarzschild case and therefore remains untouched.
\end{itemize}
Now we can run $\tt CalibratorGH.nb$ to asset an evaluation on the sampling intervals. Its outputs tell us that the scattering problem is well posed within the considered energies and parameter values. In fact, in the first plot the geometric potential is correctly going to $0$ in the asymptotic regions and the oscillatory behavior of the function $Z_s$ is well represented in both the boundary values of the considered energies (In the case of high energies it is sufficient to enlarge the plot enough to notice it). From the plots of the $Z_s$ functions one can also notice that the far-away interval on which sample and fit $Z_s$ is well defined since in such interval the function oscillates enough to ensure a good sampling.  If this is not the case one needs to change the sampling intervals accordingly to what described in Sec. \ref{Calib}.

\noindent We now implement the same steps on the main file, run it and obtain the desired result.

\section{Conclusion}
Black holes continue to be an extraordinary window into the interplay between gravity, quantum mechanics, and thermodynamics, providing fertile ground for advancing our understanding of fundamental physics. In this work, we have introduced $\tt GrayHawk$, the first publicly available Mathematica code for generating GBFs for general spherically symmetric and asymptotically flat black holes. By leveraging robust numerical techniques and the aid of Chandrasekhar transformations, our tool overcomes the inherent limitations of semi-analytical approaches, such as the WKB approximation, offering a highly accurate and versatile framework for studying black hole emission spectra.

\noindent $\tt GrayHawk$ provides a systematic method for solving the scattering problem associated with black hole perturbations. Its numerical precision and flexibility allow for accurate computations of GBFs across a broad range of black hole parameters, perturbing field spins (spins 0, 1, 2, and \(1/2\)), and energy regimes. This capability is essential for producing reliable results in scenarios where approximate methods may falter. Importantly, the degree of approximation in $\tt GrayHawk$ can be extended to any desired level, ensuring its applicability to diverse theoretical and observational investigations.  

\noindent The primary application of $\tt GrayHawk$ is to serve as a practical and accessible tool for generating GBFs with greater accuracy than traditional semi-analytical methods. Its implementation offers the scientific community a robust platform for exploring key areas of black hole physics, such as Hawking radiation, black hole evaporation, and constraints on modified BHs. By bridging the gap between theoretical predictions and numerical precision, $\tt GrayHawk$ fosters deeper insights into the rich and complex physics of black holes, particularly in addressing open questions related to primordial black hole dark matter, quantum gravity, and beyond-the-Standard-Model physics. Moreover, $\tt GrayHawk$ has already demonstrated its practical relevance by contributing to the development of the forthcoming release of $\tt BlackHawk$, namely $\tt BlackHawk\;v3.0$ \cite{BHv3}.

\noindent Future extensions of $\tt GrayHawk$ will include support for new pre-compiled metrics, expanded functionality to address current limitations (such as challenges associated with the Ghosh-Culetu-Simpson-Visser black hole metric), and the incorporation of fields arising from BSM scenarios. For instance, the inclusion of the spin-\(3/2\) gravitino field, described by the Rarita-Schwinger equation, represents an exciting avenue for future development.  

\noindent In conclusion, $\tt GrayHawk$ provides a transformative step forward in the study of black hole quantum properties, addressing long-standing limitations in the computation of GBFs and enabling precise predictions that are vital for both theoretical advancements and observational tests. By equipping researchers with this versatile tool, we aim to accelerate progress in unlocking the mysteries of black holes and their role in the broader context of cosmology and fundamental physics.

\section*{Acknowledgements}
I am particularly grateful to Sunny Vagnozzi for the precious discussions.
MC acknowledges support from the Istituto Nazionale di Fisica Nucleare (INFN) through the Commissione Scientifica Nazionale 4 (CSN4) Iniziativa Specifica ``Quantum Fields in Gravity, Cosmology and Black Holes'' (FLAG). M.C. acknowledges support from the University of Trento and the Provincia Autonoma di Trento (PAT, Autonomous Province of Trento) through the UniTrento Internal Call for Research 2023 grant ``Searching for Dark Energy off the beaten track'' (DARKTRACK, grant agreement no.\ E63C22000500003).

\end{document}